\documentclass{optica-article}

\journal{opticajournal} 

\articletype{Research Article}

\usepackage{lineno}
\linenumbers 
\usepackage[normalem]{ulem}
\usepackage{slashbox}

\begin{document}
\nolinenumbers
\title{Physical mode analysis of multimode cascaded nonlinear processes in strongly-coupled waveguides}

\author{L. Xia,\authormark{1,*} P.J.M. van der Slot,\authormark{1} C. Toebes,\authormark{2} and K.-J. Boller\authormark{1,3}}

\address{\authormark{1}Laser Physics and Nonlinear Optics Group, Applied Nanophotonics, Faculty of Science and Technology, MESA+ Institute, University of Twente, P.O. Box 217, 7500 AE Enschede, the Netherlands\\
\authormark{2}Adaptieve Quantum Optica Group, Applied Nanophotonics, Faculty of Science and Technology, MESA+ Institute, University of Twente, P.O. Box 217, 7500 AE Enschede, the Netherlands\\
\authormark{3}Optical Technologies Group, Institute of Applied Physics, University of Münster, Corrensstraße 2, 48149 Münster, Germany\\}

\email{\authormark{*}l.xia@utwente.nl} 


\begin{abstract*} 
We experimentally investigate on-chip control and analysis of spatially multimode nonlinear interactions in silicon nitride waveguide circuits. Using widely different dispersion of transverse supermodes in a strongly-coupled dual-core waveguide section, and using integrated pairs of input and output single-mode waveguides, we enable controlled excitation of nonlinear processes in multiple supermodes, while a basic physical mode decomposition aids the identification of parallel and cascaded processes. Pumping with ultrashort pulses at 1.5-$\mu$m wavelength (around 195-THz light frequency), we observe simultaneous dual-supermode, near-infrared supercontinuum generation having different spectral widths, in parallel with third-harmonic generation at around 515 nm (582 THz). Cascaded four-wave mixing with supercontinuum components upconverts the third-harmonic radiation toward a set of four shorter blue wavelengths emitted in the range between 485 and 450 nm (617 to 661 THz). The approach taken here, i.e., using chip-integrated spatial multiplexing and demultiplexing for excitation and analysis of broadband transverse nonlinear conversion, can be an advanced tool for  better understanding and control in multimode nonlinear optics, such as for extending frequency conversion to wider spectral ranges via extra phase matching paths.

\end{abstract*}

\section{Introduction}

Spatially multimode nonlinear optics represents one of the richest frontiers in nonlinear physics. Offering ultrafast spatiotemporal dynamics in an adjustable number of dimensions, the field inspires novel applications and improves fundamental understanding ~\cite{WrightLogan2022,WrightPhysicsofhighlymultimode2022}. Examples are pulse energy upscaling via a larger mode area in multi-core fibers~\cite{fang2012multiwatt}, or suppression and enhancement of Brillouin scattering, Raman scattering and spectral broadening~\cite{chen2023mitigating,tzang2018adaptive, krutova2024supercontinuum}. Other examples address extending spectral range for telecom or spectroscopic applications, by exploiting extra phase matching opportunities and cascading of nonlinear processes with an increased number of transverse modes, such as for enabling spatiotemporal mode-locking of lasers~\cite{qiu2024spectral,zeng2024wavelength} or for photon-pair generation~\cite{zhu2024tailoring}.  Interestingly, multimode nonlinear interactions can also convert fully disordered multimode input beams into a near-single-mode output, so-called beam cleaning~\cite{krupa2017spatial,dupiol2018interplay}. 

These examples are highly intriguing, however, coping with the increased complexity of multi-modal interactions requires additional means of controlling the optical input and analyzing the output. Typically, it becomes required to excite a multimode system with a known spatial mode superposition and to analyze also the modal superposition of the output, before internal nonlinear interactions can be traced.

So far, controlling the excitation of transverse mode nonlinear interactions has mostly been investigated using  bulk-optical input coupling to fibers, i.e., via tilting, translating or shaping~\cite{cao2022shaping} a free-space input beam ~\cite{Hu2003TunableCore,tzang2018adaptive}. A general drawback of bulk coupling is that the selectivity in exciting single transverse modes, or a specific superposition of these, is limited by free-space diffraction. This is especially important with tightly guided modes. Also, oblique beam injection or shaping is alignment sensitive, adding to perturbation and drift sensitivity of fiber modes~\cite{dejdar2023characterization}. Similarly, the spatial analysis of the modal content using bulk optics is affected by diffraction and alignment. Furthermore, using numerical methods to convert, e.g., camera recordings of complex field distributions~\cite{ji2023mode} into a superposition of transverse modes can be limited in resolution or be ambiguous due to lacking phase information~\cite{litvin2014doughnut,fiddy2013legacies}.

The named problems can effectively be eluded by using integrated optical waveguide circuits for transverse mode conversion. Here the intrinsic sub-wavelength stability of chip-integrated optical path lengths can provide alignment-insensitive and phase-stable excitation of transverse input modes as well as physical decomposition of output modes. These advantages have inspired theoretical investigations, e.g., to employ the intensity dependence of the refractive index for all-optical switching, in directional waveguide couplers~\cite{jensen1982nonlinear} or via transverse mode conversion~\cite{hellwig2015ultrafast}.

Recently, there are first experimental demonstrations of transverse nonlinear conversion using integrated spatial-mode multiplexing and demultiplexing for controlling on-chip four-wave mixing and Brillouin scattering ~\cite{kittlaus2017chip,signorini2019silicon, liu2021circulator}.  However, these experiments used two or more input frequencies involving low-power continuous-wave lasers in order to restrict the output to a small set of frequencies with narrow-linewidth in the 1.5-$\mu$m telecom range. So far, there is no investigation of transverse mode nonlinear generation of light forming a spectrally broadband output, with on-chip control of transverse mode excitation and physical decomposition.



Here we present the first broadband transverse-mode near-infrared to visible generation of light using on-chip conversion of pump pulses injected into a well-defined superposition of multiple transverse modes. Broadband radiation generated in a strongly-coupled dual-core waveguide structure with near-infrared (NIR) femtosecond pump pulses is physically decomposed on chip into single-core waveguide modes for a spatio-spectral analysis. Using high-power ultrashort input pulses in the near-infrared, we demonstrate cascaded nonlinear processes among multiple transverse modes enabled by the dual-core waveguide structure. Next to two different supercontinuum spectra, we observe third-harmonic generation in the green spectral range, which initiates four-wave mixing (FWM) of infrared supercontinuum radiation providing output in the blue wavelength range. A further phase-matched FWM contributes to four discrete blue wavelengths. These transverse multimode-generated frequency-conversion processes in the experiment, facilitated and unraveled via integrated on-chip input control and physical mode decomposition, respectively, can advance better understanding and control of multimode nonlinear optics. Such processes enable wider frequency conversion through unlocking extra phase-matching conditions, generating light at different frequency ranges of interest. 


\section{Controlled excitation and mode analysis using a dual-core waveguide}
\label{principle}

\begin{figure}[b!]
\centering
\includegraphics[width=0.8\linewidth]{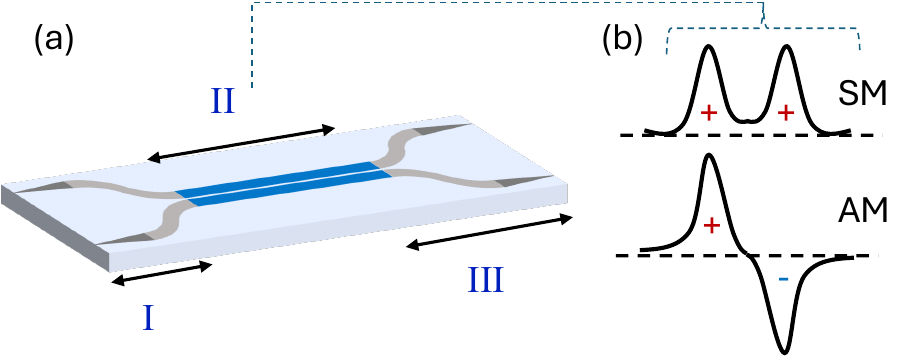}
\caption{(a) Schematic diagram for a waveguide structure with controlled excitation and analysis for multimode nonlinear optics. The device consists of three sections: (I) an input section for controlled excitation with optical pulses entering from the left into the circuit, (II) a section for multimode nonlinear optical interactions and (III) a section of physical decomposition for output analysis. (b) Schematic view of the transverse electric field distribution in the dual-core section II of the two lowest-order transverse eigenmodes (supermodes). One is a symmetric supermode (SM), and the other is an anti-symmetric supermode (AM).}
\label{fig:waveguide principle}
\end{figure}

To enable broadband multimode nonlinear generation of light in its simplest form using controlled excitation and analysis, we use an integrated waveguide structure as schematically shown in Fig.~\ref{fig:waveguide principle}~(a). The device can be divided into three sections having different functions: (I) serves for a controlled excitation of transverse modes, (II) is designed for providing nonlinear interactions, and (III) enables a physical decomposition of generated light for analysis. 

Section II is the central part of the device where multimode nonlinear optical (NLO) processes take place.  Here we use a strongly-coupled dual-core waveguide serving two purposes. First, when optically fed from the dual-pronged input section I, it allows for an efficient (i.e. mode-matched) and controlled excitation of the two lowest-order transverse field distributions, also called supermodes~\cite{Burns1988}. Figure~\ref{fig:waveguide principle}~(b) depicts schematically the two lowest-order supermode field distributions that can be excited, one being symmetric (SM) and the other being anti-symmetric (AM). The core separation of the dual-core structure allows for mode dispersion engineering beyond what is possible with a single-core overmoded waveguide~\cite{Xia23}, such as to generate two different supercontinuum spectra in parallel by simultaneously implementing normal and anomalous dispersion. Similarly, providing a large difference in supermode dispersion can offer a wider range of phase matching options, to enable cascading of four-wave mixing processes toward shorter wavelength generation.

To describe section I, where two single-mode input waveguides (input prongs) approach each other gradually, the two individual mode fields at the entrance to section I evolve gradually into the lowest-order supermodes of section II. This follows the working principle of a standard directional coupler ~\cite{Ramadan1998ADC, Mrejen2015ADC}, where two equally-dimensioned cores approach adiabatically to ensure that, at the entrance to section II, only two lowest-order supermodes become noticeably excited. More specifically, if light is injected only into one of the input prongs, as used in our experiments, this results in a well-controlled transverse mode excitation. Quantitatively, the symmetric and anti-symmetric supermodes become excited with equal amplitudes and equal or opposite relative phase. Single-prong injection of an optical field thus renders well-defined dual-spatial-mode pumping of transverse-mode nonlinear processes in section II.

The last section III enables physical mode decomposition where the reverse process compared to section I takes place. Here, light nonlinearly generated in transverse supermodes is converted into two single waveguide field distributions, which can be measured separately or in superposition. Here, two types of measurements can be performed that distinguish between two possible spatial symmetries of newly generated light. If, for instance, section II generates light solely in a spatially symmetric mode (SM), the two field distributions at the exit prongs will be in phase, and will have the same spectral envelope and strength. Therefore constructive interference is found, when superimposing the fields from the two output prongs, showing the the same spectral envelope as measurements at single prongs. In contrast, if section II generates light only in the anti-symmetric mode (AM), the output fields from the exit prongs have the same spectral content and strength again, however, they are emitted with opposite phase. As a result, superimposing the two single-prong outputs yields a field at or near zero, due to destructive interference. If light becomes generated in both supermodes, i.e., in an AM-SM superposition, the output from the two prongs will be largely different from each other due to interference between supermodes. 

In this work we use equal-amplitude excitation of the symmetric and anti-symmetric supermodes with high-power pump pulses. To identify the various nonlinear optical processes occurring in section II, we analyze the spectrally broadened output with two types of measurements behind section III, i.e., we record and compare spectra recorded behind single prongs vs. spectra of superimposed light from prongs. The spatial symmetry information gained thereby is used, together with spectral information and phase-matching calculations to analyze the presence of various parallel and cascaded nonlinear optical processes enabled by a strongly coupled dual-core waveguide structure.

\section{Methods}
\subsection{Waveguide design}
\begin{figure}[t!]
\centering
\includegraphics[width=\linewidth]{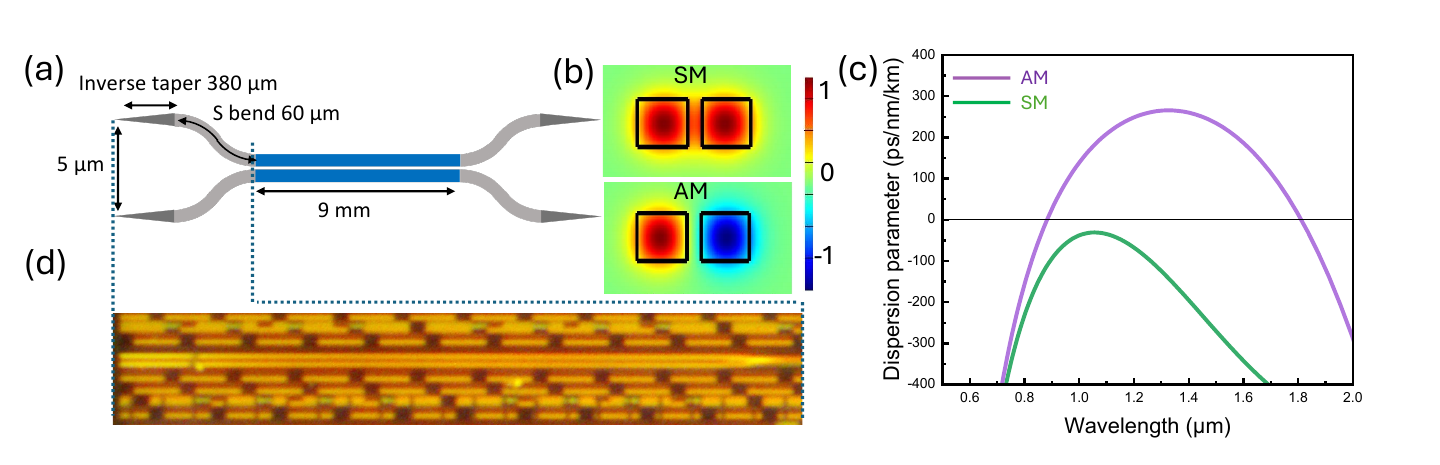}
\caption{(a) Top-view schematic of the sample. The prongs are equipped with 380~$\mu$m-long inverse tapers providing a mode-field diameter (MFD) of 2~$\mu$m at the facets. The part colored in blue is the strongly-coupled dual-core waveguide section with a small gap of 300~nm between the two cores, each 1-$\mu$m wide and 0.8-$\mu$m thick. (b) Calculated normalized field distributions ($E_\textrm{x}$ component) of the symmetric supermode (SM) and the anti-symmetric supermode (AM) supported by the strongly-coupled dual-core section at 1550-nm wavelength. The color code indicates the sign and strength of the electric field distributions, normalized to their maximum. (c) Group velocity dispersion (GVD) profiles. The SM (green curve) possesses all-normal dispersion and the AM (violet curve) possesses anomalous dispersion. (d) Microscopic top-view of the prong structure of the sample taken using the Amscope MU2003-BI camera.}
\label{waveguideDimensions}
\end{figure}

Using the Si$_{3}$N$_{4}$ platform, we converted the schematic shown in Fig.~\ref{fig:waveguide principle} into a design of a strongly-coupled dual-core waveguide structure featuring double-pronged input and output waveguides (see Fig.~\ref{waveguideDimensions}~(a) for a top-view). For a high coupling efficiency of the pump pulse at the input facet, the prongs are designed having 380-$\mu$m long inverse tapers for providing a mode-field diameter (MFD) of 2~$\mu$m. The 5-$\mu$m geometrical separation of the input prongs allows for separately exciting a single prong, to provide simultaneous and equal-amplitude excitation of the SM and the AM in the central part of the structure. The 60-$\mu$m long S-bend forming the prongs has a 1-$\mu$m wide and 0.8-$\mu$m thick cross-section designed for negligible bending loss at the nominal pump wavelength of 1550~nm. The dispersive and nonlinear characteristic lengths at 1550-nm wavelength of the input prongs are calculated to be about 1000~$\mu$m and 150~$\mu$m for a typical pulse energy of 1~nJ, respectively. With these values, dispersive and nonlinear effects are expected to be small if not negligible in the input prongs.

The multimode NLO interaction takes place in the 9-mm long, strongly-coupled dual-core waveguide section (colored in blue). To simplify the discussion of phase-matching for distinguishing between spatial modes participating in nonlinear processes, and to enable presence and cascading of various nonlinear processes for shorter-wavelength generation, we designed the initially excited two supermodes for offering a maximum difference in dispersion. This is achieved with a small gap of 300~nm between the two cores, each 1~$\mu$m wide and 0.8~$\mu$m thick. The normalized transverse field distributions ($E_x$ component) of the SM and the AM at the pump wavelength of 1550~nm, and the corresponding group velocity dispersion (GVD) profiles are calculated as depicted in figures~\ref{waveguideDimensions}~(b) and (c), respectively, using a finite element 2D mode solver (Ansys Lumerical FDE). The color code indicates the strength and sign of the electric field distributions. As shown in Fig.~\ref{waveguideDimensions} (c), these two transverse supermodes offer opposite signs of the GVD with the SM and the AM providing all-normal GVD (green curve) and strong anomalous GVD (violet curve), respectively~\cite{Xia23}.

To enable measuring the light field from the two single output prongs, both separately and in superposition, requires a compromise between sufficient separation and proximity of prongs. We fulfill these requirements with choosing a 5-$\mu$m separation, considering that a lensed fiber with 2-$\mu$m mode field diameter (MFD) is used to collect the light from a single prong separately, whereas a larger-area, 10-$\mu$m MFD, single-mode standard fiber is used to collect the superposition of the light field distributions from both prongs. Figure~\ref{waveguideDimensions} (d) gives a microscopic top-view of the prong structure.

\subsection{Experimental set-up}
\begin{figure}[t!]
\centering
\includegraphics[width=\linewidth]{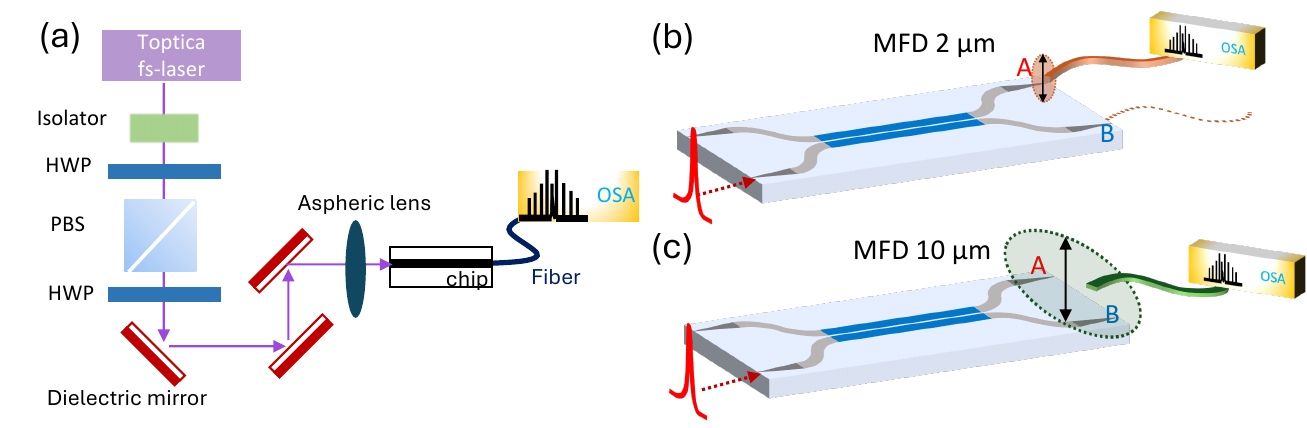}
\caption{(a) Schematic experimental set-up. The laser source operates at a center wavelength of 1544-nm with a 3-dB bandwidth of 13~nm, a pulse duration of 165~fs, and a repetition rate of 80~MHz. An optical isolator is used to protect the high-power laser source from unintended back reflections. The laser is set at full power, and the pulse power is varied outside the laser via the combination of a half-wave plate (HWP) and a polarizing beam splitter (PBS). The laser beam is coupled into a single input prong using an aspheric lens with a coupling loss of about 1~dB. Fibers are used to collect output light from the waveguides for spectral analysis using an optical spectrum analyzer (OSA). (b) First detection scheme using a small-MFD, lensed fiber for measuring output A or B separately at a time. (c) Second detection scheme using a large-MFD fiber to measure the sum of the optical fields of output A and B.}
\label{fig:setup}
\end{figure}

The experimental set-up used to investigate multimode NLO processes is schematically shown in Fig.~\ref{fig:setup} (a). Ultrashort pump pulses generated by a laser (Toptica FF Ultra 1560) with a center wavelength of 1544~nm (194.2~THz) and 165-fs pulse duration travel through an isolator, a half-wave plate (HWP), a polarizing beam splitter (PBS) and another HWP sequentially, to adjust the pulse energy external to the laser. Next, several dielectric mirrors steer the laser beam to the waveguide input facet. Here, an aspheric lens (Thorlabs C330 TMD-C, f=~3.1~mm at 1550-nm wavelength) is used for focusing the laser beam into one of the prongs, leading to a simultaneous and equal excitation of the symmetric and anti-symmetric supermodes (SM and AM) at the start of the dual-core waveguide. Output light is collected using fibers for spectral analysis. Spectral measurements are done using an optical spectrum analyzer (OSA, ANDO AQ6315A) with an optical resolution bandwidth of 50~pm in the infrared, and a second OSA (OceanInsight FLMT08730 with a 10-$\mu$m input slit) for measuring spectra in the visible. For all spectral measurements, the alignment of the input and output detection is carried out as follows. Both the input lens and detection fiber are aligned for highest transmission at pump wavelength at low pulse energy when there is no significant nonlinear conversion. Top-view pictures were taken using a microscope (Olympus SZ-CTV) and an RGB camera (iPhone 12), unless otherwise specified.

To enable physical mode decomposition, we use two detection schemes at the output. The first is shown in Fig.~\ref{fig:setup} (b) where a lensed fiber with a small mode field diameter (MFD~2~$\mu$m) is placed behind one of the output prongs at a time, to independently measure output A or B. This corresponds to observing the interference between the light generated in supermodes in a single prong, as explained in section~\ref{principle}. The second scheme shown in Fig.~\ref{fig:setup} (c) uses a large-area, MFD~10~$\mu$m, single-mode fiber (SM28), collecting and superimposing the output from both prongs. This corresponds to separately measuring the light generated only in the symmetric supermodes, as the light propagating in the anti-symmetric supermodes will not couple into the single-mode fiber. 

To measure the far-field intensity patterns of the output light, the fiber detection is replaced with a collimating, aspheric lens (Thorlabs C330TMD-C, f=~3.1~mm at 1550-nm wavelength) and a camera (Amscope MU2003-BI).

\section{Results and discussion}
\begin{figure}[b!]
\centering
\includegraphics[width=1\linewidth]{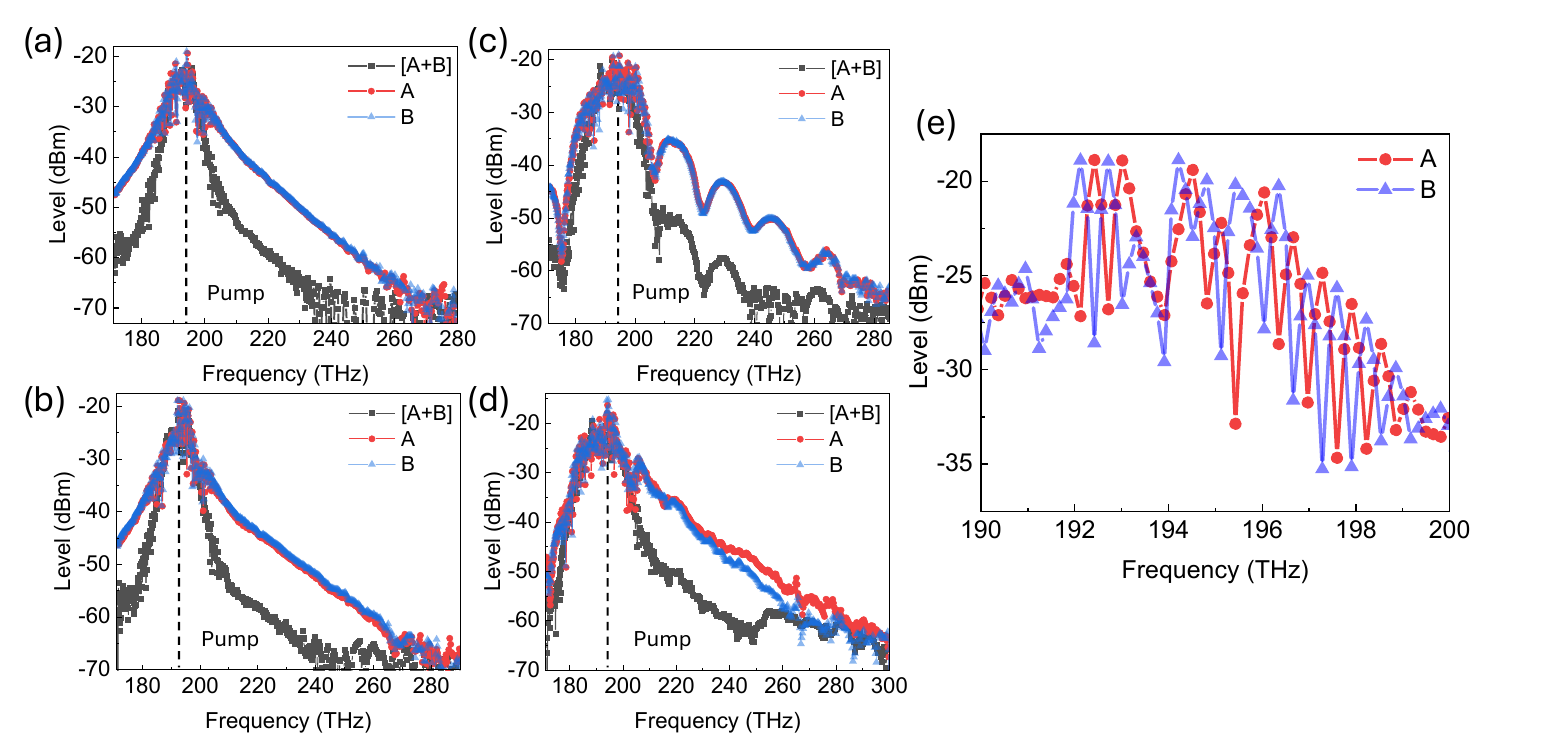}
\caption{Infrared supercontinuum output spectra from A, B and [A+B] when pumping with (a) 792~pJ, (b) 951~pJ, (c) 1.43~nJ and (d) 3.17~nJ pulse energies, respectively. (e) Spectrum in a range of frequencies closer to the pump frequency at a pump pulse energy of 951~pJ.}
\label{IRspectra}
\end{figure}

\subsection{Near-infrared broadening}
\label{NIR broadening}
The third-order susceptibility of the waveguide facilitates self-phase modulation and possible supercontinuum generation (SCG). The type of SCG is set by the GVD~\cite{GVDeffect}, and, given the two different GVD profiles (Fig.~\ref{waveguideDimensions} (c)), we expect that different supercontinuum (SC) spectra are generated by the pump light in the SM and AM supermodes. Figure~\ref{IRspectra} shows the optical spectra measured at prong A (red), B (blue) and the sum [A+B] (black) for 4 different input pulse energies of (a) 792~pJ, (b) 951~pJ, (c) 1.43~nJ and (d) 3.17~nJ. 

As [A+B] only measures light in symmetric supermodes, the black spectra in figures~\ref{IRspectra} (a)-(c) are identified with SCG in the SM. Because of the all-normal GVD in that mode, we expect the spectral width of this SC to be smaller than light generated in the AM, which has anomalous-GVD. Indeed, the black SC spectra have a smaller spectral width than the red and blue SC spectra. As a result, for frequencies sufficiently far from the pump, the SCG in the SM is weak and, there, the SCG is dominantly in the AM, which is confirmed by the spectra measured in prongs A (red) and B (blue) separately. As there is no difference in the two measured spectra, we expect only a single supermode to be present at these wavelengths, and the SCG is identified to be in the AM. Figure~\ref{IRspectra} (d), which is measured for a pulse energy of 3.17~nJ, forms an exception where SCG in the SM is comparable to that in the AM at frequencies much higher than the pump frequency. This means that at these frequencies we have the two modes simultaneously present at entrance of prongs A and B. Indeed, we now observe a difference in spectra recorded at prongs A and B separately, which is due to interference between the two supermodes. The same is true for frequencies closer to the pump frequency where SCG in both SM and AM is strong and the measurements at prongs A and B are again different (see Fig.~\ref{IRspectra} (e)), and oscillating out of phase due to the SM and AM being in- and out of phase versus the light frequency. 

These measurements show that we can physically measure and identify SCG in the SM and AM. When both supermodes are simultaneously present, the three signals (A, B and [A+B]), when calibrated, allow to determine the relative strength for each supermode. This demonstrates the feasibility of physical mode decomposition.

\subsection{Visible light generation}

\begin{figure}[hb!]
\centering
\includegraphics[width=0.85\linewidth]{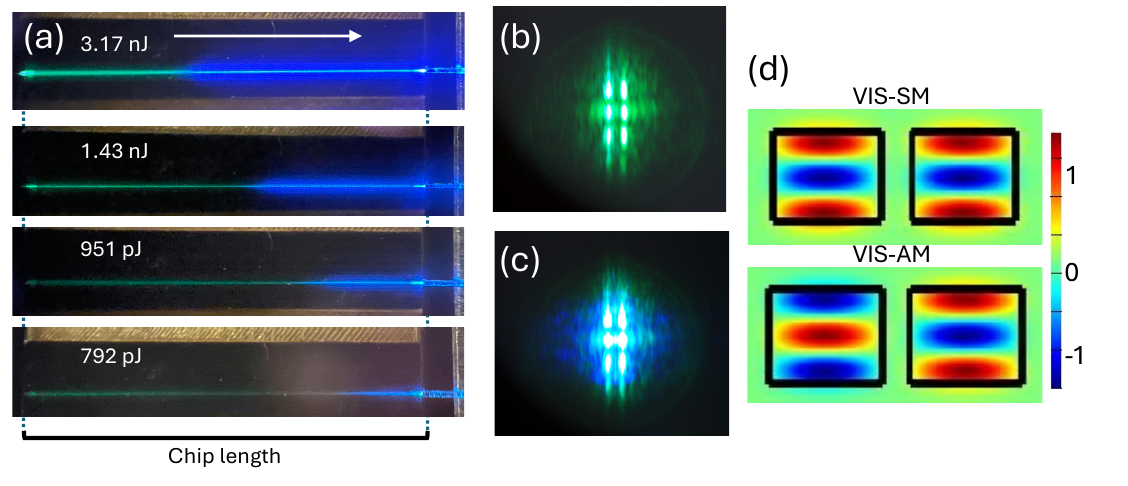}
\caption{(a) Top-view images of visible light generation along the axis of the waveguide at various pump pulse energies. The arrow indicates the light propagation direction. (b) Green and (c) blue far-field intensity patterns. (d) Calculated normalized $E_\textrm{x}$ field distributions of visible light-generated in the higher-order symmetric supermode (VIS-SM) and the anti-symmetric supermode (VIS-AM) in the strongly-coupled dual-core section. The color code indicates the sign and strength of the electric field distributions.}
\label{GreenBlueTopview}
\end{figure}

The dual-core structure in our work enables not only SCG in different transverse modes, as seen in section~\ref{NIR broadening}, but also cascaded inter-modal THG and FWM. This can be seen as the generation of green and blue light shown in Fig.~\ref{GreenBlueTopview} (a), the detailed discussion of which is given in the following sections. We note that the visible light is only observable for pump pulse energies of about 792~pJ and above. Figures~\ref{GreenBlueTopview} (b) and (c) present the far-field patterns measured in superposition of the light from prongs A and B. Using numerical wavefront propagation and calculation, we verified that the calculated spatial eigenmodes, i.e., visible light-generated higher-order symmetric and anti-symmetric modes (VIS-SM and VIS-AM) for the dual-core waveguide, as shown in Fig.~\ref{GreenBlueTopview} (d), indeed produce the measured far-field patterns, when propagated to the output facets of the prongs. Note that the far-field patterns identify only the order of the transverse modes but not its symmetry. To determine the symmetry, we again apply the physical decomposition approach, as used in the section~\ref{NIR broadening}. 

\subsubsection{Third-harmonic generation}

\begin{figure}[hb!]
\centering
\includegraphics[width=1\linewidth]{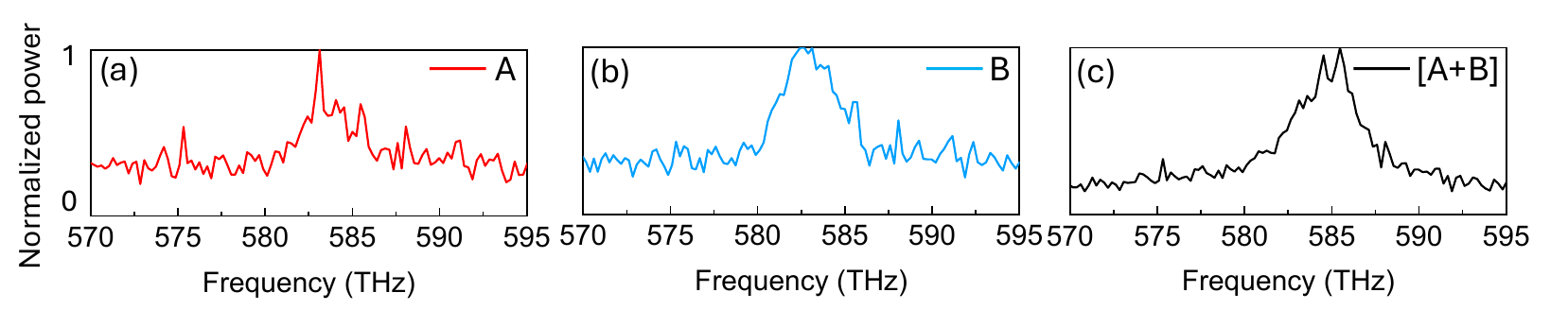}
\caption{Normalized green output spectra vs light frequency collected from (a) A, (b) B and (c) [A+B] when pumping with 792~pJ pump energy.}
\label{Green spectra}
\end{figure}

Figure~\ref{GreenBlueTopview}~(a) shows top-view images of visible light generation along the axis of the waveguide for four different pulse energies ranging from 792~pJ to 3.17~nJ. It can be seen that the green light is generated almost from the beginning of the dual-core waveguide, while the intensity increases with increasing pump pulse energy. We used the lowest pulse energy, i.e., 792~pJ for which mainly green light is generated to identify the NLO process. Figures~\ref{Green spectra} (a)-(c)  show the output spectra in the green spectral range versus light frequency collected from output A, output B and output [A+B], respectively, normalized to their maximum. We note that measuring absolute values is difficult because of different fibers used in measuring A/B (with a small-MFD lensed fiber) and [A+B] (with a large-MFD single-mode fiber), and would also require multi-octave spanning calibration. Relative differences can still be used to determine modal content.
As observed in figures~\ref{Green spectra} (a) and (b), the frequency of the green light is centered at 582.5~THz, which is at three times the pump frequency at 194.2~THz, indicating a third-harmonic generation (THG) process. We note that Fig.~\ref{Green spectra} (c) shows a slightly different center frequency of 585~THz, but also this value can be addressed to THG, generated by the secondary maximum in the spectrum of the pump laser. The optical processes are further investigated via the phase mismatch $\Delta k$ given by
 \begin{equation} \label{equation 1}
\Delta k = \frac{2\pi}{c}(f_\textrm{THG} n_\textrm{THG} - 3f_{0} n_{0}),
\end{equation} 
where $f_\textrm{THG}$ is the frequency of the THG wave, $n_\textrm{THG}$ is the refractive index of the THG wave, $f_{0}$ is the frequency of the fundamental wave, and $n_{0}$ is the corresponding effective refractive index of the dual-core waveguide. Given the THG wave at 582~THz and the fundamental wave at 194~THz, by checking various mode combinations with the THG wave in the VIS-SM or VIS-AM, and with the fundamental wave in the SM or the AM, none of them fulfills ideal phase-matching with $\Delta k=0$. Therefore, it can be concluded that the green light is generated in a non-phase-matched THG process. 
We find that the higher-order transverse modes shown in Fig.~\ref{GreenBlueTopview} (d) have the lowest phase mismatch of about $\Delta kL\approx600\pi$, which explains the observed far-field intensity pattern in Fig.~\ref{GreenBlueTopview}~(b). Here, $L\approx 9$~mm is the length of the dual-core waveguide.

According to the physical decomposition, [A+B] only measures light in symmetric supermodes, while output prongs A and B measure both symmetric and anti-symmetric modes. Although the center wavelength of the light in the SM is slightly different from that in figures~\ref{Green spectra} (a) and (b), there is still spectral overlap. Thus, if THG light is also generated in the AM, we expect some difference in the spectrum measured at A and B. Indeed, figures~\ref{Green spectra} (a) and (b) show a small difference in the spectra, confirming the presence of both supermodes. In addition, as the pump light is present in both the SM and AM,  the THG wave is likely to be generated in both the VIS-SM and VIS-AM.

\subsubsection{Green-to-blue upconversion}
\label{NIR-green-FWM}
\begin{figure}[t!]
\centering
\includegraphics[width=1\linewidth]{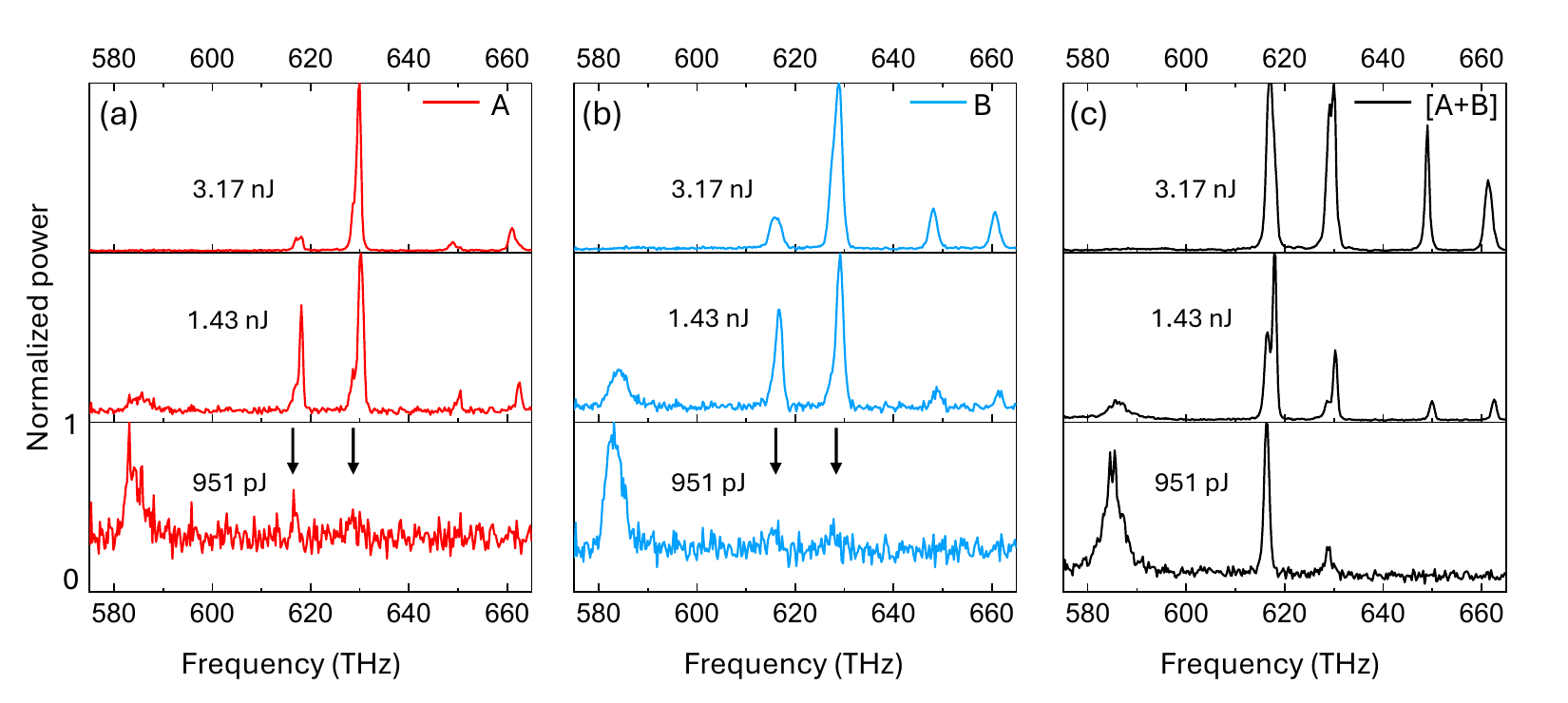}
\caption{Normalized visible output spectra vs light frequency collected from (a) A, (b) B and (c) [A+B] when pumping with different pump energy levels. }
\label{bluespectra}
\end{figure}

We now proceed to analyze the corresponding NLO process generating the blue light, as seen in Fig.~\ref{GreenBlueTopview} (a). The figure shows that, as the pump pulse energy increases, the intensities of both the green and blue emission increases as well, and that the onset of blue light generation appears at progressively shorter interaction lengths. Furthermore, the far-field pattern of the blue light is highly similar with that of the green light (figures~\ref{GreenBlueTopview} (b) and (c)), from which we conclude that the blue light is generated in the same spatial mode as the green light. All these observations seem to indicate that the green light promotes the generation of blue light. To investigate whether a phase-matched FWM process exists that upconverts the third-harmonic green radiation into the blue spectral range, we consider energy and phase-matching conservation.

The spectra of the visible light have been measured for pump pulse energies ranging from 951~pJ to 3.17~nJ and are shown as normalized spectra in Fig.~\ref{bluespectra}. For pumping with 951~pJ, blue light appears at two frequencies, 617~THz and 630~THz, as observed in the output [A+B], one of which has a spectral intensity comparable with that of the green light. According to the physical mode decomposition, this light is generated at least partially in a symmetric mode, VIS-SM. The same two frequencies are also present when the spectrum is measured separately at A (Fig.~\ref{bluespectra} (a)) and at B (Fig.~\ref{bluespectra} (b)), albeit barely above the noise, as indicated by the arrows. The A and B signals being both small, while the [A+B] signal being clearly visible is unexpected as the presence of the VIS-SM in [A+B] should at least provide a signal in A or B if not in both. We find this to be reproducible in the experiment. When the pulse energy is increased to 1.43~nJ, the spectra measured at A, B and [A+B] show four distinct frequencies, i.e., additional blue light is generated at two somewhat higher frequencies, at 648~THz and 661~THz. Here, all blue spectral components become dominant over the green light. These four blue spectral components remain present when increasing the pump pulse energy further, e.g., for 3.17~nJ (Fig.~\ref{bluespectra}). 
By applying the physical decomposition rules to the measurements at the pump pulse energy of 3.17 ~nJ, i.e., comparing the relative strength of the peaks in the spectra at A, B and [A+B], we conclude that the blue light at 630~THz and 661~THz is dominantly present in the anti-symmetric spatial mode (VIS-AM), while the blue light at 617~THz and 648~THz is dominantly present in the symmetric spatial mode (VIS-SM).

Given the four blue frequencies 617~THz, 630~THz, 648~THz and 661~THz, as the only viable NLO process we consider phase-matched blue-to-green upconversion by FWM, i.e., where the named frequencies can be generated with the green light as one of the pump waves. Specifically, this upconversion process contains two pump waves,  $f_\textrm{g}$ and $f_1$, to generate two other frequencies, $f_\textrm{b}$ and $f_2$, where energy conservation requires that
\begin{equation} \label{equation 2}
f_\textrm{b} - f_\textrm{g} = f_{1} - f_{2},
\end{equation} 
where $f_\textrm{b}$ is the frequency of the blue light, $f_\textrm{g}$ is the frequency of the green light, and $f_{1,2}$ are two frequencies to be determined to fulfill Eq.~\ref{equation 2}. Equation~\ref{equation 2} shows that the frequency difference between $f_1$ and $f_2$ is set by the observed frequency difference between the blue and green light, which is in range of 32 to 79 THz. This means that either both frequencies ($f_1$ and $f_2$) are part of the supercontinuum generated, or  that one of them is the pump laser frequency and the other is a frequency from the supercontinuum. In either case, such FWM is a cascaded process. 

The second condition to fulfill for a phase-matched FWM process is
\begin{equation} \label{equation 3}
[f_{1} n_{1}+f_\textrm{g} n_\textrm{g}] - [f_{2} n_{2}+f_\textrm{b} n_\textrm{b}] = 0,
\end{equation} 
where $n_{1}$, $n_\textrm{g}$, $n_{2}$, and $n_\textrm{b}$ are the refractive indices for the waves at frequency $f_{1}$, $f_\textrm{g}$, $f_{2}$, and $f_\textrm{b}$, respectively.
We find that simultaneously satisfying Eqs.~\ref{equation 2} and \ref{equation 3} requires the unknown frequencies $f_1$ and $f_2$ to be provided in different transverse modes, specifically, $f_1$ needs to be in the SM and $f_2$ in the AM, to provide a sufficiently different wavevector contribution (see Fig.~\ref{waveguideDimensions} (c)). Assuming nominal waveguide parameters, i.e., absence of fabrication errors, we calculate that perfect phase matching is possible for generating blue frequencies up to about 620~THz, which encompasses only the observed 617-THz emission line. As this light is in the VIS-SM, and as $f_1$ and $f_2$ are in different supermodes, it is required that the green light is in a anti-symmetric mode, VIS-AM, in order to provide a non-zero mode overlap integral~\cite{stegeman1989waveguides}. Our calculations reveal also that Eq.~\ref{equation 3} cannot be fulfilled for any of the four blue frequencies, if assuming that $f_1$ or $f_2$ is the pump frequency. Therefore, both $f_1$ and $f_2$ must originate from supercontinuum generation.

We note that drawing further conclusions on the origin of blue light generation, i.e., which and how many frequencies are expected with which spectral widths, remains difficult. One reason is that phase matching may extend also to the highest observed blue frequency of 661~THz, if including the accuracy specification for waveguide fabrication with the used silicon nitride platform (Ligentec) and the spectral width of the green light. Phase matching should also be time dependent via the time dependent power of the involved waves \cite{Agrawal_6th_Nonlinear_fiber_optics}, which has been ignored in Eq.~\ref{equation 3}. 
Furthermore, the relative generation efficiency of the observed blue frequencies should depend on the spectral variation of both (SM and AM) supercontinuum spectra, determining which pairs of $f_1$, $f_1$ maximally contribute, together with the locally varying third-harmonic intensity (at $f_\textrm{g}$). Finally, the upconversion efficiency should also depend on the temporal overlap of the involved waves, as they have different group velocities.

\subsubsection{All-blue four-wave-mixing}

Our conclusion that both frequencies, $f_{1}$ and $f_{2}$, are provided by supercontinuum in the SM and AM, respectively, actually implies that, due to the wider bandwidth of the supercontinuum spectra, various pairs of $f_{1}$ and $f_{2}$ can upconvert third-harmonic light to a wider, continuous range of blue frequencies, as discussed in section~\ref{NIR-green-FWM}. However, the measured spectra show discrete lines instead. For a pump pulse energy of 3.17~nJ, Fig.~\ref{bluespectra} shows that four discrete blue frequencies are present.
We have evaluated both energy conservation and phase matching and find that these four frequencies are mutually connected with an all-blue, phase-matched FWM process. In addition, the spatial symmetry deduced for the modes at the four blue frequencies confirms their function in providing a non-zero mode overlap integral here as well. However, revealing the origin of line emission vs broadband upconversion appears intricate. Most likely the light at 630~THz serves as one of the two pump waves in the all-blue FWM process, as it has the largest optical power. The second pump would then be the light at 648~THz, generating light at both 617~THz and 661~THz via FWM. Possibly, these two frequencies are provided by green-to-blue upconversion as discussed in section~\ref{NIR-green-FWM}, while fulfilling all-blue phase matching determines the remaining frequencies (617 and 661 THz). 


\begin{figure}[ht!]
\centering
\includegraphics[width=\linewidth]{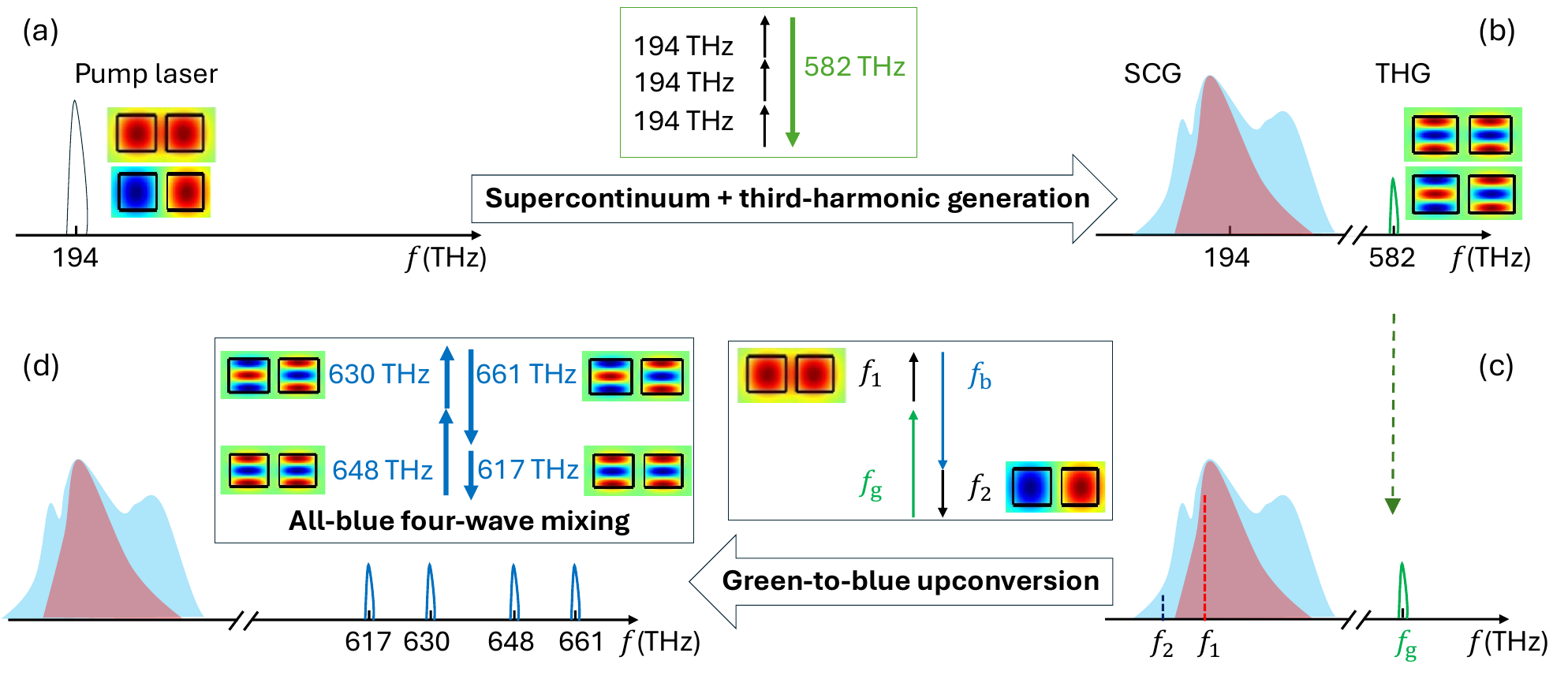}
\caption{
Schematic overview of the parallel and cascaded nonlinear processes identified in the experiment and calculated transverse mode profiles ($E_\textrm{x}$) involved. (a) Pump spectrum, (b) supercontinuum and third-harmonic spectra, (c) spectral components involved in upconversion FWM to generate blue light, and (d) four discrete blue frequencies fulfilling all-blue phase-matched FWM. 
}
\label{Overview}
\end{figure}

\section{Summary and conclusion}
We have experimentally investigated how to excite and physically decompose broadband transverse nonlinear optical processes, for exploring a path toward improved control and analysis in multimode nonlinear optics. Using a femtosecond infrared drive laser for simultaneously exciting the two lowest-order supermodes of a strongly coupled dual-core waveguide, we have identified multiple parallel and cascaded nonlinear optical transverse-mode interactions as schematically summarized in Fig.~\ref{Overview}. At the entrance to the dual-core waveguide, pump radiation at 194~THz is injected in both lowest-order supermodes, SM and AM, with equal amplitudes (see Fig.~\ref{Overview} (a)). Supercontinuum generation causes spectral broadening in the near-infrared in both supermodes with different spectral width caused by the significantly different supermode dispersion in the infrared. Simultaneously green light is generated through non-phase-matched third-harmonic generation at 582~THz in two higher-order visible supermodes, VIS-SM and VIS-AM (a$\rightarrow$b). When pumping with sufficiently high pulse energies, the presence of third-harmonic light at $f_{g}$ and frequency pairs from the supercontinuum ($f_{1}$ and $f_{2}$, Fig.~\ref{Overview} (c)) generates blue output $f_{b}$ through green-to-blue upconversion in a cascaded four-wave mixing process (c$\rightarrow$d). As four-wave mixing requires a non-zero overlap integral, the blue light is likely generated in both the VIS-SM and VIS-AM. Specifically, for green light in the VIS-SM, the blue is generated in the anti-symmetric mode, VIS-AM, and vice versa. Another cascaded four-wave mixing processes is present involving both the VIS-SM and the VIS-AM, fulfilling phase matching with non-zero mode overlap at four distinct frequencies (Fig.~\ref{Overview} (d)).

The wide spectral range of output light can be addressed to several transverse modes participating in the nonlinear light generation and to pumping at two supermodes simultaneously. In the particular system investigated here, combining integrated control of mode excitation with integrated  physical mode decomposition has assisted in analyzing the complicated sequence of multimode nonlinear interactions depicted in Fig.~\ref{Overview}, beyond what is possible solely based on energy and momentum conservation arguments. We see a potential to generate light toward UV by controlling the cascading processes via altering the waveguide circuit design to tailor phase matching paths involving multiple transverse supermodes.

We conclude that, as a prototype example, the investigated control and analysis of transverse mode nonlinear interactions may be an interesting approach for better understanding and further advancing multimode nonlinear optics. Having observed unusual multimode nonlinear cascading, the results may inspire improved numerical modelling where transverse-mode supercontinuum calculations~\cite{Poletti2008DescriptionFibers} are extended toward multi-octave spanning spectra via including mixed and cascaded nonlinearities~\cite{voumard2023simulating}. The experimental results suggest that richer multimode nonlinear dynamics with wider phase-matching opportunities may be exploited, for instance, to extend the spectral range of synchronously pumped optical parametric oscillators ~\cite{timmerkamp2024}. Other options can be generating multimode quantum states for upscaling on-chip quantum light sources~\cite{roman2021continuous}. More immediate steps in experimental investigations can be the integration of on-chip adjustability in supermode excitation, such as including tunable Mach-Zehnder waveguide couplers as proposed in ~\cite{Xia23}. A general extension can be stepwise increasing the number of coupled waveguide and thus the dimensionality toward waveguide arrays~\cite{christodoulides2003discretizing}.

\section*{Funding} This work is partially funded by the Dutch Research Council (NWO) SYNOPTIC OPTICS program, P17-24 project 1.

\section*{Acknowledgments} The authors would like to thank C. Fallnich and M. Timmerkamp for offering us to put our waveguide circuit design on their wafers.

\section*{Disclosures} The authors declare no conflicts of interest.

\bibliography{Citation}

\end{document}